\documentclass[english,aps,preprint]{revtex4}
\usepackage[T1]{fontenc}
\usepackage[latin9]{inputenc}
\usepackage{babel}

\usepackage{verbatim}
\usepackage{amsmath}
\usepackage{graphicx}
\usepackage{amssymb}
\usepackage[unicode=true, pdfusetitle,
 bookmarks=true,bookmarksnumbered=false,bookmarksopen=false,
 breaklinks=false,pdfborder={0 0 1},backref=false,colorlinks=false]
 {hyperref}

\makeatletter
\@ifundefined{textcolor}{}
{%
 \definecolor{BLACK}{gray}{0}
 \definecolor{WHITE}{gray}{1}
 \definecolor{RED}{rgb}{1,0,0}
 \definecolor{GREEN}{rgb}{0,1,0}
 \definecolor{BLUE}{rgb}{0,0,1}
 \definecolor{CYAN}{cmyk}{1,0,0,0}
 \definecolor{MAGENTA}{cmyk}{0,1,0,0}
 \definecolor{YELLOW}{cmyk}{0,0,1,0}
 }

\makeatother

\begin{document}

\preprint{BROWN-HET-1593}

\title{Punctuated eternal inflation via AdS/CFT}

\author{David A. Lowe$^{\dagger}$ and Shubho Roy$^{\dagger\natural\sharp}$ }

\email{lowe@brown.edu, sroy@het.brown.edu}

\affiliation{$^{\dagger}$Department of Physics, Brown University, Providence,
RI 02912, USA}

\affiliation{$^{\natural}$Physics Department City College of the CUNY New York,
NY 10031, USA}

\affiliation{$^{\sharp}$Department of Physics and Astronomy Lehman College of
the CUNY Bronx, NY 10468, USA}
\begin{abstract}
The work is an attempt to model a scenario of  inflation in the framework
of Anti de Sitter/Conformal Field theory (AdS/CFT) duality, a potentially
complete nonperturbative description of quantum gravity via string
theory. We look at bubble geometries with de Sitter interiors within
an ambient Schwarzschild anti-de Sitter black hole spacetime and obtain
a characterization for the states in the dual CFT on boundary of the
asymptotic AdS which code the expanding dS bubble. These can then
in turn be used to specify initial conditions for cosmology. Our scenario
naturally interprets the entropy of de Sitter space as a (logarithm
of) subspace of states of the black hole microstates. Consistency
checks are performed and a number of implications regarding cosmology
are discussed including how the key problems or paradoxes of conventional
eternal inflation are overcome.
\end{abstract}
\maketitle

\section{Introduction}

Making contact with realistic cosmology remains the fundamental challenge
for any candidate unified theory of matter and gravity. Precise observations
\cite{Riess:1998dv,Perlmutter:1999rr} indicate that the current epoch
of the acceleration of universe is driven by a very mild (in Planck
units) negative pressure, positive energy density constituent - {}``dark
energy''. Likewise there is strong evidence that the Big Bang phase
of the universe was preceded by a exponentially accelerated (inflation)
phase \cite{Guth:1980zm,Linde:1981mu} driven by a negative pressure,
positive energy fluid. In most cosmological models this inflationary
stage is brought about by a scalar field coupled to gravity, slowly
{}``rolling down'' an almost flat potential hill, and upon reaching
the bottom of the hill gives rise to the big-bang stage. In most such
models inflation is inevitably {}``future eternal'' i.e. there are
always some residual regions which keep on rapidly inflating. The
steady state picture that emerges is a fractal multiverse structure
with many {}``pocket universes'' causally separated by continually
inflating sterile regions. These {}``pocket universes'' might possess
all possible different {}``fundamental constants'' of nature (including
a cosmological constant). So in this scenario one now talks about
{}``environmental constants'' of nature instead.

(Super)string theory is the currently leading candidate for an unified
theory of fundamental interactions and should be compatible with realistic
cosmology if it has to be anything more than an attractive toy model.
It is great news that string theory appears to have an enormous {}``landscape''
of vacua including many long-lived metastable states with positive
cosmological constants \cite{Kachru:2003aw} forming a quasi-continuum
\cite{Bousso:2000xa,Feng:2000if,Giddings:2001yu,Susskind:2003kw}
which are highly relevant for realistic cosmology. These metastable
phases ultimately decay to neighboring stable anti-de Sitter phases
via {}``rolling down the hill'' and quantum mechanical tunneling
and this is exactly what one needs to attempt a fundamental physics
model of the (inflationary) multiverse. So it is important to develop
tools in the fundamental theory which describe such transitions.

\begin{figure}
\includegraphics[scale=0.7]{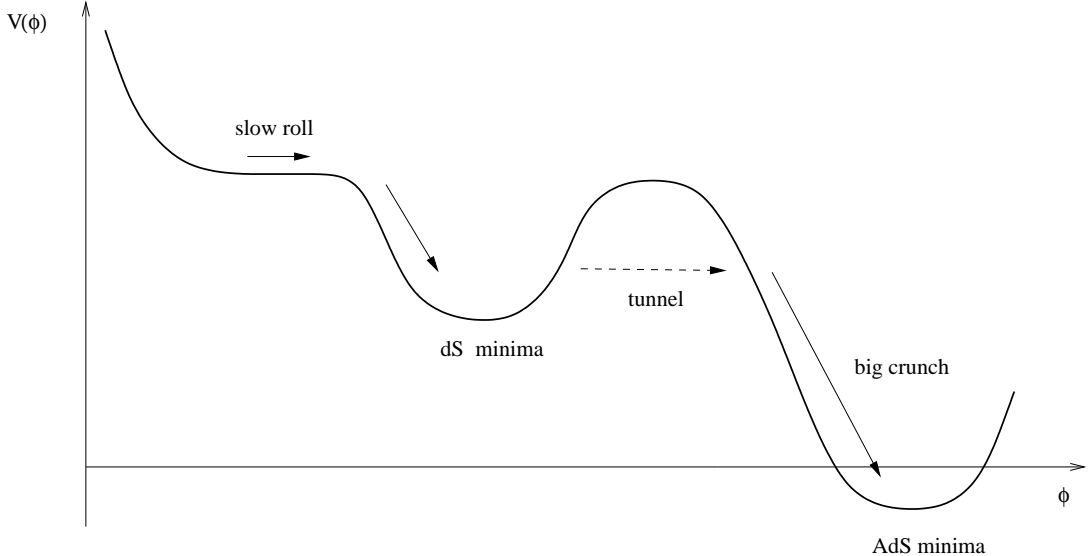}\caption{The potential landscape showing the dS metastable and AdS stable vacua.
The arrows indicate the time evolution of a possible worldline.\label{fig:The-potential-landscape}}

\end{figure}

The AdS/CFT formulation of String theory \cite{Aharony:1999ti} is
a potentially nonperturbative definition of superstring theory in
asymptotically AdS spaces and is an perfect setting to embed the string
theory landscape. The neighborhood of the landscape that we study
is summarized in figure \ref{fig:The-potential-landscape}. %

The first step within this approach was taken in \cite{Alberghi:1999kd}
which looked at classical AdS-Reissner-Nordstrom/dS domain wall type
geometries (an inflating dS bubble interior patched to an asymptotic
AdS black hole spacetime) in the thin wall approximation. Classical
bubble spacetimes without charge were further investigated in \cite{Freivogel:2005qh}
and subsequently in \cite{Lowe:2007ek}. The latter has an added attractive
feature that the mysterious Gibbons-Hawking entropy of de Sitter space
could be identified as a (logarithm of) subspace of black hole microstates.
The aim of the present paper is to build on the work of \cite{Lowe:2007ek}
by attempting to identify the CFT states which code these inflating
states in a thermal ensemble, and further explore the consequences
of this scenario.

The paper is organized as follows. In section 2, we review the semiclassical
consistency condition of \cite{Lowe:2007ek} which arises from interpreting
de Sitter entropy within a unitary framework for quantum gravity.
This amounts to the condition that the number of states available
must be bounded below by the Gibbons-Hawking entropy if states corresponding
to a truly semiclassical region of de Sitter space exist. . Then in
section 3, we review the classical bubble spacetimes. Keeping in mind
the dual CFT as the underlying quantum description we are forced to
rule out the whole class of time-symmetric bubble solutions based
on the consistency condition. In particular, this rules out tunneling
solutions of the type described in \cite{Farhi:1989yr}. In section
4, we focus on the CFT picture of the time-asymmetric solutions compatible
with the aforementioned consistency condition and their signatures
in the dual CFT. We model the effect of the radiation reflecting off
the shell by reducing the problem to that of a moving mirror in a
black hole background. In section 5 we discuss consistency checks
of such a scenario. In section 6 we draw various conclusions regarding
the implications for such a picture for cosmology and how several
problems/paradoxes that plague the eternal inflation scenario - namely
the measure problem, the youngness paradox and the Boltzmann brains
paradox are overcome in this setting.

\section{Quantum De Sitter state\label{sec:Quantum-De-Sitter}}

In \cite{Lowe:2007ek} it had been advocated that the Gibbons-Hawking
entropy of dS space can be correctly interpreted as the logarithm
of the dimension of quantum gravity state (Hilbert) space. This proposition
was based on numerical observations of the entropy of the present
universe, dominated by a contribution from the cosmic microwave background
(CMB) photons. We review this rationale briefly here once again. The
entropy of CMB photons see, for example, \cite{Frampton:2008mw})
is,

\[
S_{CMB}\sim10^{88}\]
with wavelength, $\lambda_{CMB}\sim1$ mm, while the Gibbons-Hawking
entropy of dS space with the present day cosmological constant is,

\[
S_{GH}\sim10^{123}.\]
So with a field theory cut off with about $\epsilon=1$ TeV, the total
number of states in effective field theory associated with CMB photons
is $(\lambda_{CMB}\epsilon)^{3}S_{CMB}\sim10^{138}$, way too large
to be accounted by the de Sitter entropy. 

From past experience with AdS/CFT we know that the notion of a cut
off in the bulk effective field theory (EFT) is complex and in particular
dependent on the respective quantum (CFT) state, unlike the conventional
uniform cut off in usual effective field theories \cite{Hamilton:2006fh,Hamilton:2006az,Hamilton:2005ju,Lowe:2008ra}.
So an EFT in the bulk with an adaptive step size cutoff which allocates
up to $10^{35}$ states per CMB photon would saturate the Gibbons-Hawking
entropy of dS space with the present day cosmological constant. Hence
we are justified in saying that Gibbons-Hawking de Sitter entropy
does count (logarithm of) the number of possible states in EFT, albeit
with a clever cut-off \cite{Lowe:2007ek}.

AdS/CFT also taught us that generic states in the quantum theory (CFT)
have no nice bulk space time interpretation. But in addition to just
a smooth space time geometry (metric) once also requires large families
of observables to reproduce the entire span of semiclassical physics
on this smooth geometry. This observation coupled with the above numerical
bound on the entropy of the universe leads us to the following hypothesis
for the de Sitter state in a quantum theory of gravity:
\begin{itemize}
\item \textbf{Semiclassical Consistency Hypothesis (SCH): }\textbf{\emph{\it{The set of microstates representing a semiclassical de Sitter region must number at least }
$e^{S_{GH}}$ .}}
\end{itemize}

\section{Inflation in $\mbox{AdS/CFT}$: $\mbox{dS/SAdS}$ domain walls\label{sec:Inflation-in-:}}

\begin{figure}
\includegraphics{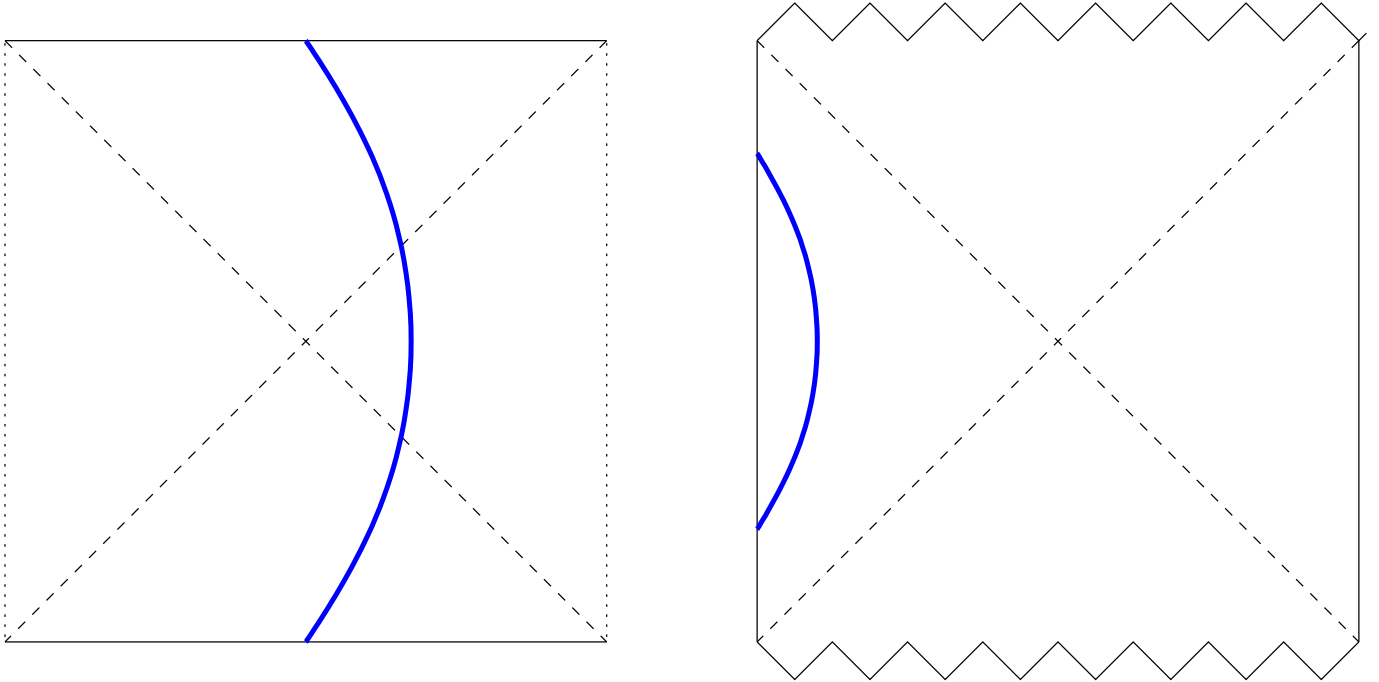}\caption{A Time symmetric bubble trajectory: The left diagram is dS spacetime
and the right diagram is SAdS spacetime. The blue curve is the bubble
trajectory. The domain wall geometry is made up of the region to the
left side of the blue curve in the dS spacetime diagram and to the
right side of the blue curve of the AdS spacetime diagram.\label{fig:A-Time-symmetric}}

\end{figure}

Let us briefly review the geometric set up which can be found in great
detail in \cite{Blau:1986cw,Alberghi:1999kd,Freivogel:2005qh}. We
have a domain wall geometry by patching together a (interior) dS-region
with cosmological constant (c.c.) $\Lambda_{dS}=\frac{3}{l_{dS}^{2}}$
and an (exterior) AdS-Schwarzschild geometry with c.c set to $\Lambda_{AdS}=\frac{3}{l_{AdS}^{2}}$
and hole mass $M$ across a thin spherical shell (parametrized $r=R(\tau)$
where $\tau$ is the proper time of the shell. The world volume metric
of the shell is then given by the equation,

\[
ds^{2}=-d\tau^{2}+R(\tau^{2})d\Omega^{2}\,.\]
The geometry is specified by the following metric and the coordinate
ranges,

\[
ds^{2}=\begin{cases}
-f_{dS}(r)dt_{dS}^{2}+\frac{dr^{2}}{f_{dS}(r)}+r^{2}d\Omega^{2}, & f_{dS}(r)=1-\frac{r^{2}}{l_{dS}^{2}},\qquad r<R(\tau)\:(\mbox{inside)}\\
-f_{SAdS}(r)dt_{SAdS}^{2}+\frac{dr^{2}}{f_{SAdS}(r)}+r^{2}d\Omega^{2}, & f_{SAdS}(r)=1-\frac{2GM}{r}+\frac{r^{2}}{l_{AdS}^{2}},\: r>R(\tau)\,\mbox{(outside)}\,.\end{cases}\]
Here $r$ is a global coordinate and the times inside and outside
the bubble are related by equating the invariant distance along the
shell from inside and outside.

\begin{equation}
\left(-f_{dS}(r)dt_{dS}^{2}+\frac{dr^{2}}{f_{dS}(r)}\right)_{r=R(\tau)}=-d\tau^{2}=\left(-f_{SAdS}(r)dt_{SAdS}^{2}+\frac{dr^{2}}{f_{SAdS}(r)}\right)_{r=R(\tau)}\,.\label{eq:locals}\end{equation}
The consistent junction conditions relate the discontinuity of the
extrinsic curvature across the shell to the stress energy tensor of
the shell (defining $\kappa=4\pi G\sigma$, with $\sigma$ the surface
energy density of the shell),%
{}

\[
\sqrt{\dot{R}^{2}+f_{SAdS}(r)}-\sqrt{\dot{R}^{2}+f_{dS}(r)}=\kappa R\]
which determine the shell trajectory (substituting $r=R(\tau)$) can
be rewritten

\begin{equation}
\dot{r}^{2}+V_{eff}(r)=0\label{eq:reom}\end{equation}
so the trajectory is that of a zero net energy particle moving in
potential,

\begin{equation}
V_{eff}(r)=-\frac{\left(\frac{1}{l_{dS}^{2}}+\kappa^{2}-\frac{1}{l_{AdS}^{2}}\right)^{2}+\frac{4}{l_{dS}^{2}l_{AdS}^{2}}}{4\kappa^{2}}r^{2}+1+\frac{GM\left(\frac{1}{l_{dS}^{2}}+\frac{1}{l_{AdS}^{2}}-\kappa^{2}\right)}{\kappa^{2}}\frac{1}{r}-\frac{G^{2}M^{2}}{\kappa^{2}r^{4}}\label{eq:veff}\end{equation}
in one space dimension. The solutions can be broadly classified into
two categories - one when the parameters are such that the potential
has a maximum that is positive and when it is negative. In the first
case the zero energy particle is reflected off the barrier and the
bubble traces out a time symmetric trajectory, in some choice of coordinates,
while in the second case the particle rolls over to the top of the
potential hill to the other side and the bubble traces out a time
asymmetric trajectory , figure \ref{fig:A-Time-symmetric}. 

As shown in \cite{Freivogel:2005qh} for all the time symmetric cases:\[
r_{bh}<r_{dS}\]
 i.e. the black hole horizon radius is less than the de Sitter horizon
length and this implies,

\[
S_{GH}>S_{BH}\,,\]
with $S_{BH}$ the Bekenstein-Hawking entropy of the black hole. 

Now let us consider a black hole such that $r_{bh}>l_{AdS}$ so that
the black hole is well-described by the canonical ensemble in the
CFT. The available microstates, $e^{S_{BH}}$, must exceed the microstates
of the internal dS bubble, $e^{S_{GH}}$, according to the SCH. There
are simply not enough microstates in the dual CFT to constitute the
whole range of semiclassical phenomena in the interior of the bulk
dS bubble. So the SCH rules out all such time symmetric bubble configurations
in the quantum picture. In particular no such false vacuum dS bubbles
can be formed by quantum mechanical tunneling from AdS spacetime.
Thus we see in the AdS/CFT framework, the type of tunneling events
considered in \cite{Farhi:1989yr,Fischler:1989se} do not occur. Once
any kind of local tunneling process allows arbitrarily large regions
of semiclassical spacetime to be formed, we quickly violate unitarity,
so this is an important new constraint that arises from the CFT viewpoint,
that would not be seen via a straightforward semiclassical gravity
analysis.

\begin{figure}
\includegraphics{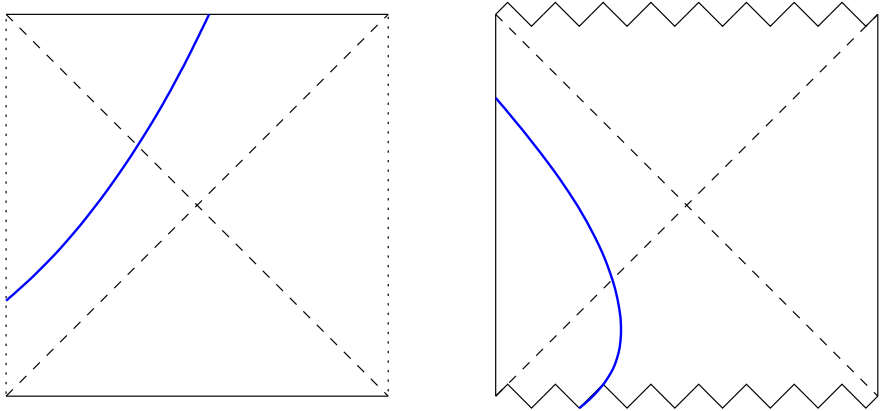}\caption{Time asymmetric trajectory : The interior dS region is shown on the
left, and the exterior SAdS region is shown on the right. The domain
wall geometry is made up of the regions to the left of the blue curve
in the dS diagram and to the right of the SAdS diagram joined across
the bubble wall (blue curve). \label{fig:Time-asymmetric-trajectory}}

\end{figure}

We are left with the time asymmetric situation where the bubble emerges
from the past singularity $R(\tau=0)=0$ and keeps on expanding indefinitely
i.e. $r(\tau\rightarrow\infty)\rightarrow\infty$ to simulate an eternally
inflating bubble of de Sitter space inside an asymptotic AdS space.
The conformal diagram of such a geometry is shown in figure \ref{fig:Time-asymmetric-trajectory}. 

This can be ensured if the outward normal points to increasing $r$
when $r\rightarrow\infty$ i.e. by having $\beta_{dS}>0$ which gives

\[
\kappa^{2}>\frac{1}{l_{dS}^{2}}+\frac{1}{l_{AdS}^{2}}\,.\]
For these time asymmetric cases Gibbons-Hawking dS entropy can be
accounted for by the black hole microstates (i.e. one can arrange
for $r_{bh}>l_{dS}$) by having

\[
\left(2GMl_{AdS}^{2}\right)^{1/3}>l_{dS}.\]

After taking into account back reaction as was shown in \cite{Lowe:2007ek},
the left asymptotic boundary of the bubble exterior ceases to exist
as this region is unstable to perturbations and a singularity develops.
In the bubble interior, any worldline eventually tunnels back into
the stable AdS vacuum, but the tunneling portion of the bubble undergoes
a big crunch \cite{Abbott:1985kr}.

\begin{figure}
\includegraphics{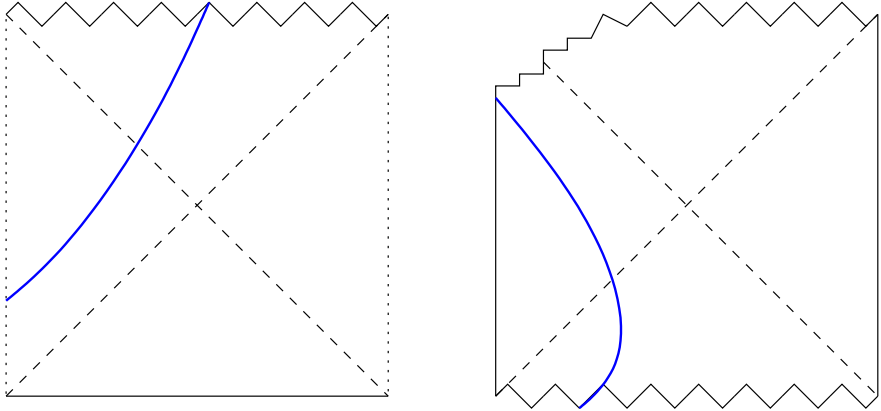}

\caption{Penrose diagrams after taking into account perturbations. The left
diagram shows the bubble trajectory in dS spacetime. The region to
the right of the bubble wall is excised. Eventually any given worldline
tunnels back into an AdS region, leading to a big crunch. The diagram
on the right shows the trajectory through AdS spacetime. The region
to the left of the wall is excised. Perturbations lead to a Cauchy
horizon with a null singularity meeting the bubble wall at infinity.\label{fig:Penrose-diagrams-after}}

\end{figure}

\section{THE CFT PICTURE}

\subsection{Schwarzschild anti-de Sitter black hole}

\begin{figure}
\includegraphics[scale=0.7]{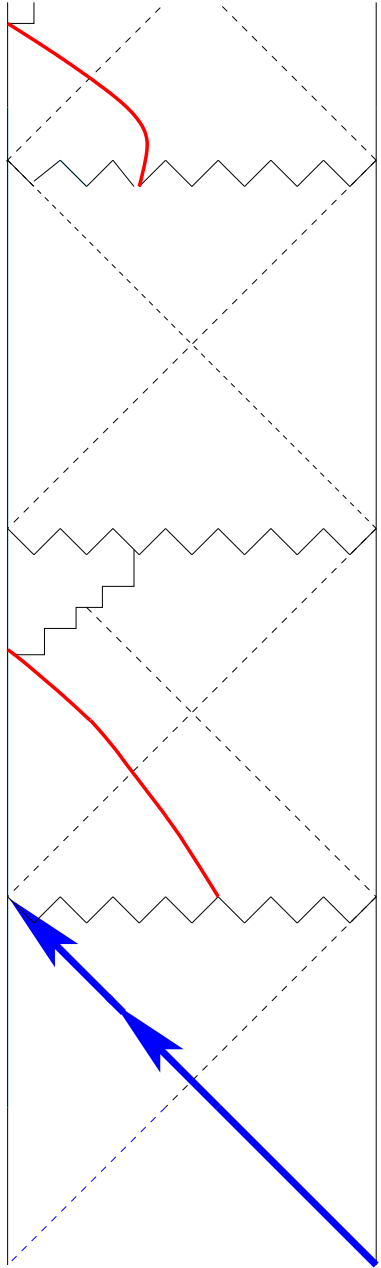}

\caption{Punctuated Eternal Inflation: Starting with empty AdS space, dual
to the vacuum state of the CFT, a light-like shell, shown by the blue
line, collapses into a black hole. A bubble with dS interior, shown
by the red curve, appears out of the spacelike singularity. The left
side of the red line i.e. the bubble wall is a piece of dS space.
This sequence repeats, following the quasiperiodic Poincare recurrences
in the dual gauge theory. \label{fig:Punctuated-Eternal-Inflation: Figure 4}}

\end{figure}

The basic setup we consider is shown in figure \ref{fig:Punctuated-Eternal-Inflation: Figure 4}.
A large black hole is formed from the collapse of a null shell sent
in from infinity into empty $AdS_{4}$ spacetime. This process is
expected to be described by a pure state within a single conformal
field theory. Under time evolution, the state is thermalized, and
late-time correlation functions will be well-approximated by those
in the canonical ensemble, at temperature corresponding to the Hawking
temperature of the black hole, or alternatively in the microcanonical
ensemble at fixed energy. 

The conformal field theory is unitary and can be viewed as living
on the left conformal boundary $\mathbb{R}\times S_{2}$. Since the
spatial directions are compact the CFT entropy in these statistical
ensembles will be finite and given by the Bekenstein-Hawking entropy
of the black hole $S_{BH}$. For timescales of order the Poincare
recurrence time $e^{S_{BH}}$, the correlators will deviate wildly
from the thermal correlators \cite{Birmingham:2002ph,Barbon:2004ce,Barbon:2005jr,Solodukhin:2005qy,Kleban:2004rx,Festuccia:2006sa},
but ultimately will settle down again to quasi-thermal correlators.
This is represented in figure \ref{fig:Punctuated-Eternal-Inflation: Figure 4}
by singularities separating well-behaved semiclassical regions with
a SAdS black hole exterior. 

Note that some of these regions apparently have two disconnected boundary
regions at infinity \cite{Maldacena:2001kr} (such as the second region
from the top in figure \ref{fig:Punctuated-Eternal-Inflation: Figure 4}).
However, in line with the setup in the previous paragraphs, these
asymptotic regions are not to be viewed as completely independent
of each other, but rather as representing analytic continuation of
modes with finite energy from a single exterior region, dual to states
in a single CFT. Ultimately these bulk geometries are approximate
descriptions of the time evolution of a pure state in a single CFT.

In general states emerging from the white hole singularities in figure
\ref{fig:Punctuated-Eternal-Inflation: Figure 4} will contain all
excitations present in the thermal ensemble. In particular, the bubble
solutions of section \ref{sec:Inflation-in-:} will appear with nonvanishing
probability. Since the Poincare recurrences go on forever, we have
an infinite number of times to produce bubbles with realistic cosmological
interiors, including inflation, while at any given time only a finite
number of CFT states need be considered. These transitions populate
the string theory landscape in a $timelike$ manner instead of a $spacelike$
manner as in standard eternal inflation \cite{Vilenkin:1983xq,Steinhardt:1982kg,Linde:1986fc,Linde:1986fd,Goncharov:1987ir,Linde:1993xx,Linde:1994gy,GarciaBellido:1994ci}.
We refer to this scenario as \emph{punctuated eternal inflation. }As
we will see, many of the attractive features of eternal inflation
survive in this picture, while at the same time, problems of the cosmological
measure and many other issues are ameliorated. 

\subsection{Bubble states in the CFT\label{sub:Bubble-states-in}}

It is helpful to begin by first considering modes of a scalar field
with finite global energy, and the relevant vacuum state. Let us review
Unruh's construction \cite{Unruh:1976db} adapted to the eternal SAdS
geometry. The geometry includes two asymptotic regions related via
analytic continuation. Any finite frequency Kruskal modes can be continued
analytically across the horizon and used to provide a complete basis
in the full SAdS manifold. Likewise, Rindler modes can be patched
together to define global modes. For the Rindler modes, positive frequencies
are defined with respect to asymptotic timelike killing vectors, and
positive frequency for global modes is defined with respect to the
null killing vector at the horizon when the metric is written in Kruskal-like
coordinate\cite{Unruh:1976db}. %
\begin{figure}
\includegraphics[scale=0.7]{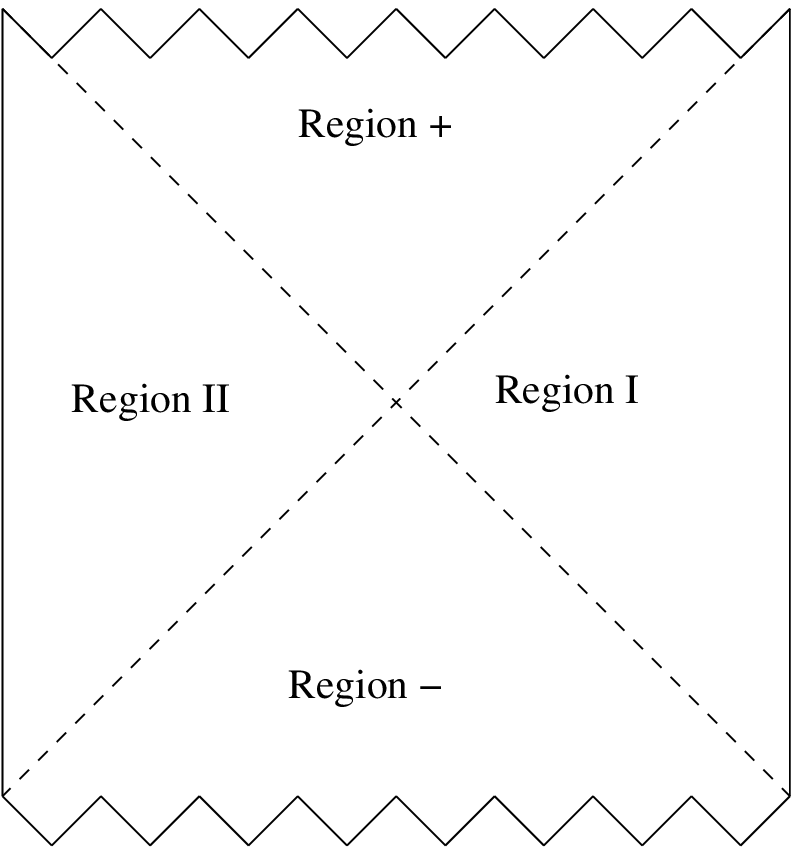}\caption{The two asymptotic regions of SAdS}

\end{figure}

The two types of vacuum state are related by

\[
|0\rangle_{Kruskal}=\prod_{\omega}\exp(e^{-4\pi GM\omega}\, b_{1\omega}^{\dagger}b_{2\omega}^{\dagger})|0\rangle_{1}\times|0\rangle_{2}\]
where $1,2$ indicate the left/right asymptotic regions of SAdS and
$b_{1/2}$ ($b_{1/2}^{\dagger}$) are the respective Rindler annihilation
(creation) operators. The Kruskal annihilation operator can expressed
in terms of the creation and annihilation operators on the left and
right regions as follows,

\begin{equation}
a_{\omega}^{\dagger Kruskal}=b_{1\omega}^{\dagger}+e^{-4\pi GM\omega}b_{2\omega}\,.\label{kruskal-mode}\end{equation}
Thus we see that a finite energy mode well-localized in one asymptotic
region will necessarily have an exponentially suppressed component
in the other asymptotic region. These finite frequency bulk modes
may then be mapped to operators in the conformal field theory via
the standard bulk to boundary map of \cite{Gubser:1998bc,Witten:1998qj},
generalized to the eternal black hole background.

\begin{figure}
\includegraphics[scale=0.7]{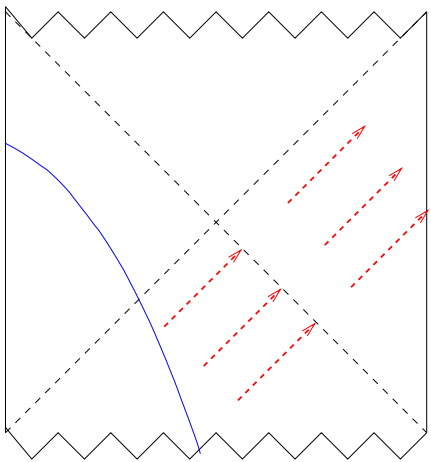}\caption{Radiation coming\label{fig:Radiation-coming-off} off the bubble walls}

\end{figure}

Now we have so far made the convenient approximation that the bubble
wall is thin. The essential details of the construction are not expected
to change if we generalize to consider walls of finite thickness,
that might be built from finite frequency modes. Moreover, we expect
operators creating and annihilating bubbles to obey a similar equation
to \eqref{kruskal-mode}. This is good news for representing such
operators in the CFT, since they may be reconstructed using the positive
and negative frequency modes in only the right asymptotic region,
which may then be mapped into the CFT. In addition to this leading
order effect, further details of the bubble state can be extracted
from the radiation coming off the bubble wall to the left asymptotic
region (depicted by the red dashed arrows in the figure \ref{fig:Radiation-coming-off}).

To study the effects of the radiation coming off the bubble we model
the bubble wall as a perfect mirror i.e. impose purely reflecting
boundary conditions for the (scalar) field propagating in the SAdS
background and restrict ourselves spherical symmetry. The computation
is sketched in Appendix \ref{sec:Appendix-B}. Further specification
of the bubble would require the particular details of the metastable
string theory vacuum giving rise to the potential shown in figure
\ref{fig:The-potential-landscape}. Nevertheless, for a given potential
and choice of initial state in the gravity dual, one has a well-defined
(one-to-many) mapping into the CFT.

%
{}

At this point let us collect a few facts on the spectrum of bubble
states, which will be useful in the subsequent sections. One could
compute frequencies of normalizable excitations around the bubble
geometry by solving the gravity equations of motion. However this
spectrum is expected to receive important corrections due to back-reaction
and quantum effects. The CFT enables us to take these effects into
account. Even without all the details of the CFT dual that incorporates
a bulk potential of the form figure \ref{fig:The-potential-landscape},
some qualitative features of the spectrum can be determined. The domain
wall states considered in the present scenario are viewed as a subset
of black hole microstates. These are dual to states in the CFT$_{3}$
on the spacetime $\mathbb{R}\times S_{2}$, therefore the entire spectrum
of CFT states will be discrete. We can schematically write the microcanonical
density matrix to be,

\[
\rho(E)=\sum_{a,b,...}|\psi_{a,b,....}\rangle\langle\psi_{a,b,...}|\]
where\[
\mbox{{\bf H}}|\psi_{a,b,....}\rangle=E|\psi_{a,b,....}\rangle\]
are energy eigenstates with $discrete$ internal labels $a,b,....$
. For large total energy $E$ this can be well-approximated by the
density matrix of the canonical ensemble

\[
\rho(\beta)=\sum e^{-\beta\mbox{{\bf H}}}|\psi_{a,b,\cdots}\rangle\langle\psi_{a,b,\cdots}|\]
where $\beta$ is the black hole inverse temperature. This may then
be directly mapped into the canonical ensemble in the CFT at inverse
temperature $\beta$. This will be our basic working definition of
the ensemble of states that contains inflating bubbles of the type
shown in figure \ref{fig:Radiation-coming-off}. As we will see, this
is already sufficient to provide interesting bounds on the measures
of cosmological interest.

Of course the canonical ensemble also contains many states that do
not represent cosmological solutions of interest. One could construct
a more refined ensemble by selecting on black holes with expanding
interiors. Local bulk correlators can be constructed in these spacetimes
(including points inside the bubble) by generalizing the two-dimensional
construction of \cite{Lowe:2008ra}. Correlators of massless fields
will exhibit a characteristic powerlaw falloff with geodesic separation
inside the bubble, and this geodesic separation depends directly on
the vacuum energy inside the bubble. Using the methods of \cite{Hamilton:2005ju,Hamilton:2006az,Hamilton:2006fh}
these correlators may be mapped into integral transforms of correlators
in the CFT. This yields a straightforward, though calculationally
tedious way of selecting the vacuum energy inside the bubbles.

Another, more direct way, to select on expanding bubble states is
to use characteristic radiation coming off the bubble (as shown in
figure \ref{fig:Radiation-coming-off}) to refine the specification
of the CFT state. As mentioned above, this is explored in a simplified
model in appendix \ref{sec:Appendix-B}. On the gravity side the solution
is that of a black hole with a set of quasinormal modes excited. The
amplitudes for these modes will be determined by the bubble initial
state, and the detailed properties of the potential of figure \ref{fig:The-potential-landscape}.
This can be represented by a time-dependent density matrix in the
CFT, that will, in general, depend on the vacuum energy inside the
bubble.

\section{Consistency checks\label{sec:Consistency-checks}}

The semiclassical consistency hypothesis of section \ref{sec:Quantum-De-Sitter}
by construction produces quantum versions of de Sitter space with
a sensible semiclassical limit as $\hbar\to0$ and $S_{GH}\to\infty$.
In this section we perform some consistency checks on the approach
to the semiclassical limit, with a view to keeping $S_{GB}$, $S_{BH}$
and $\hbar$ finite to use as a physically sensible regulator for
eternal inflation. The detailed scenario we have in mind for the remainder
of the paper is shown in figure \ref{fig:The-potential-landscape},
where a worldline begins in a rapidly inflating phase, that later
settles into a metastable de Sitter vacuum, that will be modeled as
the late-time development of bubbles solutions such as those shown
in figure \ref{fig:Radiation-coming-off}. 

We begin by studying the late-time history of this worldline. From
the original work \cite{Coleman:1980aw,Abbott:1985kr} we know that
a worldline in the dS bubble tunnels back to the stable AdS vacuum.
This event is accompanied by extra excitation energy that quickly
leads to a big crunch. Related tunneling solutions within the AdS/CFT
context have recently been studied in \cite{Barbon:2010gn}. From
an estimate of the tunneling rate we can we know how long this worldline
lasts in proper time. This time is (see, for example \cite{Kachru:2003aw}
)

\begin{equation}
\tau_{decay}\sim e^{S(\phi)}e^{S_{GH}},\label{eq:decaytime}\end{equation}
where $S(\phi)<0,\;|S(\phi)|\gg1$ and is just the $O(4)$ invariant
Euclidean action for the tunneling trajectory from dS to AdS space.
This may be converted into a timescale in the CFT by mapping to Schwarzschild
time in the exterior AdS region. This calculation is performed in
appendix \ref{sec:Appendix-A}, eqn. \eqref{eq:sadstime}.

\[
t_{decay,SAdS}\approx B-\frac{Al_{AdS}^{2}}{l_{dS}}e^{-e^{cS_{GH}}},\;0<c<1\]
where $A,B$ are constants defined in appendix \eqref{sec:Appendix-A}.
Thus the decay time is typically shorter than a Planck time in SAdS
coordinates, from the moment when the bubble wall reaches the null
singularity in the exterior region, as shown in figure \eqref{fig:Penrose-diagrams-after}.
For such short timescales we should not trust the semiclassical gravity
description, though the CFT description is valid. We therefore conclude
that this tunneling process happens long after conventional semiclassical
physics has broken down in the dS bubble. 

The Poincare recurrence time for de Sitter space $\tau_{Poincare,\: dS}\sim e^{S_{GH}}$
is already much longer than the tunneling timescale \eqref{eq:decaytime}.
This eliminates all the issues one encounters when the decay time
exceeds the recurrence time, for instance breakdown of a classical
general relativity by reversal of the arrow of time in a semiclassical
regime as discussed in \cite{Dyson:2002pf}. 

The finest time resolution we can reasonably hope to resolve in the
bulk is \[
\Delta t_{SAdS}=l_{pl}\]
the Planck time. Translating this back into de Sitter proper time
(as shown in appendix \ref{sec:Appendix-A}) we obtain\begin{equation}
\tau_{dS}=l_{dS}\log\left(\frac{Al_{AdS}^{2}}{l_{pl}l_{dS}}\right).\label{eq:plscale}\end{equation}
As we will later see, this timescale will typically be $l_{dS}10^{d}$
where $d$ is some number of order 1. Thus we see that in this scenario
our semiclassical de Sitter bubble survives for a far shorter than
one might have imagined based on tunneling probabilities, before the
discreteness of the CFT makes its presence felt. Nevertheless, as
we will see, for reasonable parameter choices, our present cosmological
history can be embedded in this scenario.

Now let us address the question of whether for times smaller than
\eqref{eq:plscale} the CFT can accurately reproduce correlators inside
the dS bubble. This issue arises in studies of the black hole information
problem, where the CFT must encode the details of the internal state
of the black hole. As shown in \cite{Lowe:1999pk,Lowe:2006xm}, for
effective field theory in black hole backgrounds, the departure from
locality on observations made over times less than the scale in Schwarzschild
coordinates $\Delta t_{SAdS}\sim S_{BH}\beta_{H}$ is less than the
scale $e^{-S_{BH}}$ (for example the amplitude of Hawking emission)
following information theoretic arguments first invoked in \cite{Page:1993wv}.
Recall that the AdS$_{5}$/CFT$_{4}$ map implies $S_{BH}\propto N^{2}$
and hence $e^{-S_{BH}}\sim\frac{1}{e^{N^{2}}}$ represents effects
which are non-perturbative in $1/N$ in the dual gauge theory. For
the $d=3$ case the dual CFT is expected to be a generalization of
the CFT studied in \cite{Aharony:2008ug}. There the CFT was realized
as a Chern-Simons theory with Chern-Simons level number $k$ playing
the role of gauge coupling, $g_{YM}^{2}$.

The observer in the dS bubble will only be able to measure, for example,
energy with an intrinsic error of $1/l_{dS}$ given the above discussion.
This is already a much larger error than the error potentially allowed
from a construction of bulk amplitudes from the exact CFT. Therefore
we conclude there are no immediate obstructions to reproducing semiclassical
physics inside the bubble from the exact CFT description, for proper
times less than \eqref{eq:plscale}.

\section{Implications for cosmology}

In this section we examine a number of standard issues in cosmology
within this scenario.

\subsection{Bound on the number of e-foldings of inflation}

Suppose we make a bubble with $\Lambda_{dS}=\Lambda_{today}\sim(10^{-3}eV)^{4}$
and say it developed from a GUT scale bubble excitation $\Lambda_{GUT}=(10^{16}GeV)^{4}$
following the time evolution sketched in figure \eqref{fig:The-potential-landscape}.
Let us denote the maximum number of e-foldings in the high-scale phase
by $n_{max}$. If we use the Gibbons-Hawking entropy to count states
in the different de Sitter phases, then balancing the entropy before
and after high-scale inflation, we have\begin{align}
\mbox{No. of Hubble volumes }\times S_{GUT} & \leq S_{today}\nonumber \\
n_{max} & \leq\frac{1}{3}\log\frac{\Lambda_{GUT}}{\Lambda_{today}}\nonumber \\
n_{max} & \leq86.\label{eq:nbound}\end{align}
So we see the present scenario reproduces the bound on the number
of efolds derived in \cite{Banks:2003pt}. These kinds of bounds are
discussed further in \cite{Lowe:2004zs}. 

The logic presented here is rather different to that of \cite{Banks:2003pt}.
Here we match the number of available microstates within the ensemble
of states that evolves toward a single large $\Lambda_{today}$ bubble
to derive the bound. In \cite{Banks:2003pt} they instead consider
the entropy of a fluid in de Sitter and match the entropy of the stiffest
fluid, that of a gas of black holes, in order to reach the bound \eqref{eq:nbound}.
Nevertheless it is satisfying the bound of \cite{Banks:2003pt} is
reproduced when de Sitter spacetime is embedded in a complete unitary
theory of quantum gravity, such as the AdS/CFT duality. The present
scenario offers a number of modifications and refinements to the \emph{Cosmological
Complementarity} principle presented in \cite{Banks:2001yp}.

\subsection{Natural minimization of the cosmological constant}

A fundamental puzzle in modern cosmology is smallness of the cosmological
constant compared to the scales of fundamental physics. In terms of
Hubble parameters, there is a huge hierarchy\[
\frac{H_{today}}{H_{inflation}}\propto\sqrt{\frac{\Lambda_{today}}{\Lambda_{GUT}}}\sim10^{56}\,.\]
The present scenario gives us a natural explanation for decrease of
the cosmological constant. Let us consider the ensemble of microstates
that can evolve to a single large bubble, which we will refer to as
the grand ensemble. For timescales smaller than \eqref{eq:plscale}
we treat this ensemble as a closed system. Semiclassical evolution
of the bubble implies that it keeps growing for as long as possible.
The classical singularity theorems of Hawking and Penrose \cite{Hawking1970}
imply the initial state began at high density. As the bubble expands,
the number of states is ultimately bounded by $e^{S_{BH}}$, the number
of black hole microstates in the external AdS spacetime.

We model the GUT-scale phase of inflation by a similar ensemble with
$\Lambda_{dS}=\Lambda_{GUT}$, which we refer to as the small ensemble.
With each e-folding an entire Hubble volume worth observables are
added to the small ensemble. We should view the GUT-scale region as
contained within the grand ensemble of the previous paragraph, since
ultimately it evolves toward it. Therefore we can apply the second
law of thermodynamics to the small ensemble and find that the entropy
(the log of the number of available microstates) increases as the
bubble expands. This, of course, can only be consistent if $\Lambda_{today}<\Lambda_{GUT}$.
Extending this argument by relating the cosmological entropy in a
background with time-dependent Hubble parameter \cite{Easther:1999gk}
to the microscopic entropy, one finds\[
\dot{S}\geq0\Rightarrow\dot{H}\leq0\,.\]

A closely related argument has been put forward by Strominger \cite{Strominger:2001gp}
based on a $c-theorem$ by viewing the de Sitter regions as dual to
CFTs. He proposed our universe is represented as a renormalization
group (RG) flow between two conformal fixed points where the forward
time-translation in the bulk was interpreted as an IR-to-UV RG flow
of a putative dual CFT. The Hubble parameter,

\begin{equation}
H=\frac{\dot{a}}{a}\label{eq:hubble}\end{equation}
was proposed to play the special role of the (inverse of) central
charge of the dual CFT,

\begin{equation}
c=\frac{1}{HG}\label{eq:centralcharge}\end{equation}
because for matter obeying the null energy condition, the Einstein
equation implies,

\[
\dot{H}\leq0\Rightarrow\dot{c}>0\,.\]

In the context of the present scenario, the conformal symmetry of
the de Sitter phase arises as an emergent symmetry of the subset of
black hole microstates corresponding to the expanding bubbles, over
a restricted range of timescales. Nevertheless, because the gravity
problem can be understood in terms of a RG flow, it suggests that
these de Sitter solutions do correspond to stable (or at least quasi-stable)
fixed points. It remains an open problem how to better identify these
fixed points more directly once embedded in the consistent AdS/CFT
framework.

\subsection{Unitarity and eternal inflation}

One of the features of the standard eternal inflation scenario is
that once started, there is always a region on a given spacelike slice
where high-scale inflation is occurring. Essentially baby universes
branch off in these highly quantum regions. It is difficult to imagine
how such processes can preserve unitarity in a quantum description.
If the number of microscopic degrees of freedom changes with time,
then unitarity is necessarily violated. Certainly if such branching
processes occurred locally, we can invoke the results of \cite{Banks:1983by}
that show that such violations of unitarity would lead to Planck scale
violations of energy and momentum conservation in the bulk. 

Aside from the issue of unitarity, standard eternal inflation has
the attractive feature that the probability distribution of inflating
regions should eventually reach a stationary state. This has the potential
to lead to some very interesting predictions, however as we will review
later, the infinite nature of the spacelike slices in this stationary
state has so far stymied this effort. 

Another useful feature of eternal inflation is that it provides a
natural mechanism for populating the landscape of string theory vacua.
There is always some region on a given spacelike slice in a position
to roll down to one of the stable or metastable vacua of string theory.
This phenomena then opens the door to anthropic arguments playing
a role in our understanding of the observed parameters of nature.

In the bubble scenario, unitary is no longer an issue, since the states
are realized as states in a unitary conformal field theory. As we
have seen, this regulates both the spacelike and timelike extent of
a cosmological region that may be treated semiclassically. However
as far as we have seen, there is room to describe a large expanding
universe for a large number of Hubble times. 

As we have seen, the semiclassical consistency hypothesis of section
\eqref{sec:Quantum-De-Sitter} rules out tunneling to semiclassical
expanding bubbles. Because baby universes cannot form from smooth
initial data, the problems with bulk unitarity described in \cite{Banks:1983by}
are avoided. Instead expanding bubbles only emerge from punctuation
points, i.e. the spacelike singularities of black/white holes. Unlike
standard eternal inflation, a time-independent probability distribution
does not emerge. 

However Poincare recurrences in the CFT lead to quasi-periodic recurrence.
Thus rather than in the spacelike direction, we have infinite extent
in the timelike direction. The randomness of the recurrences can provide
a mechanism to populate whatever stable and metastable vacua are present,
so this attractive feature of eternal inflation is preserved.

\subsection{Measure on the space of initial data}

Now as emphasized before we can work in a microcanonical ensemble
since the spacetime goes through many $punctuated$ phases and has
sufficient time to explore the entire phase space. So at the fundamental
level, the measure on the space of initial data should be set by ergodicity.
In particular, all bubble states at equal energy will be equally probable.
The precise probability distribution will depend on the details of
the spectrum of black hole microstates in the particular CFT dual
to the gravity background with potential shown in figure \ref{fig:The-potential-landscape}.
However a number of general properties of these distributions can
be determined simply from a knowledge of the black hole entropy, which
must bound the bubble entropy, and the interpretation of bubble entropy
described in section \ref{sec:Quantum-De-Sitter} and in \cite{Lowe:2007ek}.

For single bubbles, one could approximate the probability distribution
as \begin{equation}
P_{bubble}\left(l_{dS}\right)=\frac{e^{S_{GH}(l_{dS})-q(l_{dS},M)}}{e^{S_{BH}(M)}},\label{eq:probdist}\end{equation}
where $M$ is the fixed mass of the large AdS black hole that dominates
the microcanonical ensemble, and $q(l_{dS},M)\geq0$ represents a
kind of form factor that parameterizes the likelihood of the black
hole having such bubble solutions as microstates. Qualitatively, we
expect $\lim_{l_{dS}\to l_{pl}}q(l_{dS},M)=0$ since the formation
of small bubbles near the singularity should be a local process, independent
of $M$. However $\lim_{l_{dS}\to r_{BH}}q(l_{dS},M)\gg1$, where
$r_{BH}$ is the black hole horizon radius, reflecting the fact that
no bubble states larger than the black hole are allowed. This means
large single bubbles are more likely, with a typical size $l_{pl}<l_{dS}<r_{BH}$. 

Now let us consider the effect of multiple bubbles. For very large
bubbles, only a single bubble will fit inside the black hole, so the
question becomes whether there is more entropy in a soup of small
bubbles. Let us approximate the collection of small bubbles by a gas
of particles in thermal equilibrium with the anti-de Sitter Schwarzschild
black hole. The entropy of this gas will be bounded above by the entropy
of a gas of massless particles (which for the sake of simplicity we
can take to be conformally coupled). This one of the systems studied
in the work by Hawking and Page \cite{Hawking:1982dh}. They find
that for $M>m_{pl}^{2}l_{AdS}$ that the black hole entropy dominates
over the entropy of the massless particles. Here $m_{pl}^{2}=1/G$,
with $m_{pl}$ the Planck mass. Assuming the form factor $q(l_{dS},M)$
does not fall off very rapidly, the entropy of single large bubbles
will therefore dominate over that of multiple small bubbles, for sufficiently
large black holes. The probability distribution for different values
of $l_{dS}$ can then be well-approximated by \eqref{eq:probdist}
assuming the gravity theory has access to a full landscape of string
theory vacua with many different values of $l_{dS}$ possible, so
that $q(l_{dS},M)$ is a smooth function.

\subsection{Resolution of cosmological paradoxes}

In the present scenario time development is future eternal, and at
any given time the region along a spacelike slice undergoing interesting
cosmology (i.e. a large region in an expanding phase) is of finite
extent. It will be useful to count the number of potential observers
within a given causal diamond inside this region. To do this we compute
the entropy per Hubble volume inside the expanding bubble as a function
of time. Here we have in mind a bubble with an interior rolling down
the potential as shown in figure \ref{fig:The-potential-landscape},
so for now on we will generalize to a Robertson-Walker metric for
the interior\[
ds^{2}=-dt^{2}+a(t)^{2}(dr^{2}+d\Omega^{2})\,.\]
Denoting the densities by $\rho_{\psi},\rho_{\gamma}$ and $\rho_{\Lambda}$
for the densities of matter, radiation and cosmological constant respectively,
the Friedman equations are

\[
\dot{a}(t)-a(t)\sqrt{\frac{8\pi l_{pl}^{2}}{3}\left(\rho_{\psi}+\rho_{\gamma}+\rho_{\Lambda}\right)}=0\]

\[
\dot{\rho}_{\psi}(t)+3\frac{\dot{a}(t)}{a(t)}\rho_{\psi}(t)=0\]
\[
\dot{\rho}_{\gamma}(t)+4\frac{\dot{a}(t)}{a(t)}\rho_{\gamma}(t)=0\]

\[
\dot{\rho}_{\Lambda}(t)=0\,.\]
The entropy density for any component is\[
s_{i}=\frac{(\rho_{i}+P_{i})}{T_{i}}=\frac{(1+w_{i})\rho_{i}}{T_{i}}\]
where $P=w\rho$ is the equation of state ($w=0,\frac{1}{3},-1$ for
matter, radiation and, cosmological constant respectively). Now the
temperature of each component is given by,\[
\frac{1}{T_{i}}=\frac{\partial S_{i}}{\partial E_{i}}=\frac{\partial s_{i}}{\partial\rho_{i}}\]
and the entropy density as a function of the $\rho$ is then,\[
\frac{s_{i}}{s_{i0}}=\left(\frac{\rho_{i}}{\rho_{i0}}\right)^{\frac{1}{1+w_{i}}}\]
and similarly the energy density of each component is,\[
\frac{\rho_{i}}{\rho_{i0}}=\left(\frac{a(t)}{a_{0}}\right)^{-3(1+w_{i})}\,.\]
So the entropy per Hubble volume from matter and radiation is,\begin{equation}
S_{H}(t)=\sum_{i}s_{i}\left(\frac{1}{H}\right)^{3}=s\left(\frac{a_{0}}{\dot{a}(t)}\right)^{3}\label{eq:entropytime}\end{equation}
where $s=\sum_{i}s_{i0}$ and we sum over only matter and radiation
components, and take this as a measure of the number of potential
observers in the bubble as a function of time. In addition, there
is a small contribution from Hawking radiation associated with the
cosmological horizon

\begin{equation}
S_{\Lambda}=\frac{\pi^{2}}{15}T_{dS}^{3}\frac{1}{H_{dS}^{3}}\sim O(1)\,.\label{eq:entropylam}\end{equation}
This component is constant with time, and dominates at late times
. This leads to the issue of so-called Boltzmann Brain observers,
that we discuss momentarily. 

\begin{figure}
\includegraphics[scale=0.6]{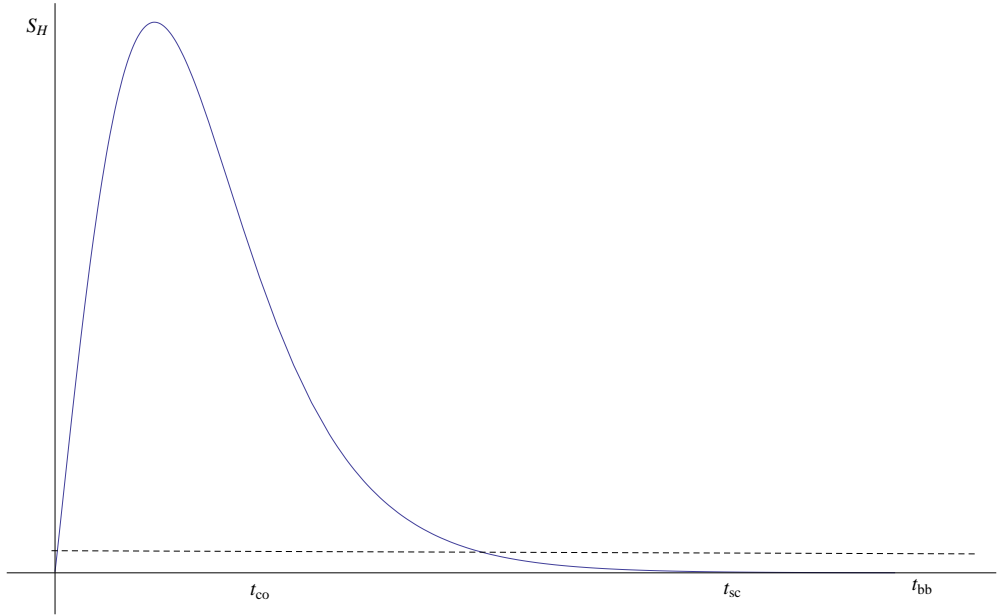}\caption{Entropy of matter and radiation per Hubble volume as a function of
time. This provides an estimate of the number of potential observers
as a function of time.\label{fig:Entropy-of-matter}}

\end{figure}

The entropies are shown in figure \ref{fig:Entropy-of-matter}. The
peak of the entropy curve is located at $\ddot{a}=\sum_{i}\left(\rho_{i}+3P_{i}\right)=0$
i.e. at \begin{equation}
\rho_{\psi}+2\rho_{\gamma}-2\rho_{\Lambda}=0\label{eq:peak}\end{equation}
which typically happens for $t<l_{dS}$ when there is a balance between
matter, radiation and cosmological constant energy densities. This
will be less than the timescale $t_{sc}$ when semiclassical physics
breaks down \eqref{eq:plscale} if we have a region where gravity
is a good description inside the bubble. Also shown on the plot is
the timescale when Boltzmann brains dominates $t_{BB}$, which will
always be much larger than the timescale \eqref{eq:plscale}. This
figure will play a key role in seeing how various cosmological paradoxes
are resolved.

\subsubsection{Cosmological measure\label{sub:Cosmological-measure}}

The theory of inflation has had many successes in terms of explaining
a number of observed phenomena within our Hubble volume, such as providing
natural explanations of the horizon problem, the flatness problem
and the spectrum of cosmic microwave background perturbations. Extending
the theory of inflation using semiclassical approaches to quantum
gravity leads to eternal inflation \cite{Vilenkin:1983xq,Steinhardt:1982kg,Linde:1986fc,Linde:1986fd,Goncharov:1987ir,Linde:1993xx,Linde:1994gy,GarciaBellido:1994ci},
which describes a multiverse of pocket universes. The goal then becomes
how to ascribe probabilities to histories through this multiverse,
i.e. the cosmological measure problem.

In eternal inflation the {}``pocket universes'' are infinite in
number. Without any regularization, probabilities, being a ratio of
two infinities, are ill-defined. Even after regularization there is
no guarantee that the answer could not depend on the regularization
method. Current attempts to define the measure tend to be coordinate
dependent and and hence ambiguous (\cite{Guth:2007ng} and references
therein). 

This issue is resolved in the present scenario: a natural regulator
that preserves unitarity is introduced. A probability measure for
different cosmologies can then be extracted by evaluating \eqref{eq:probdist}
for a given conformal field theory, which computes the likelihood
of expanding bubbles as a function of their parameters.

As emphasized in \cite{Carter:1974zz,Weinberg:1987dv} what is most
relevant for observation are probabilities conditioned on the existence
of observers. This can be incorporated into the discussion in the
following way. The density of observers can be expected to be proportional
to the entropy density of matter (and perhaps also radiation). However
entropy density alone is not a good guide, because one could have
a high density, rapidly expanding region of spacetime. In this case
the number of degrees of freedom in causal contact with each other
can remain small, so complex systems, such as observers, will not
form. 

Therefore we propose to count the amount of matter and radiation per
Hubble volume as an approximate measure of the number of degrees of
freedom in causal contact with each other. This has the advantage
of being a locally defined quantity along a given worldline. Moreover
it reduces to the volume of a causal diamond when the cosmological
constant dominates the expansion. 

The result is a world-line measure that determines the likelihood
an given observer will measure a particular set of landscape parameters,
as well as other cosmological observables. For example, if we condition
on the the observer being in an expanding region where semiclassical
gravity operates, then we conclude from \eqref{eq:probdist} and figure
\ref{fig:Entropy-of-matter} that the most likely observer will exist
at time $\tau=\tau_{peak}$.

\subsubsection{Cosmic Coincidence Problem:}

Another fundamental puzzle of cosmology is why the energy density
of dark energy is comparable to the energy of the matter today \begin{equation}
\rho_{\Lambda}\sim\rho_{Matter}\,.\label{eq:coinc}\end{equation}
This is resolved by the world-line measure described in the previous
section. A typical observer will appear near the peak of the entropy
per Hubble volume, where the densities are related by \eqref{eq:peak}. 

This computation is similar in spirit to the argument of \cite{Garriga:1999hu}.
There they generalize Weinberg's argument \cite{Weinberg:1987dv}
for the most likely cosmological constant and treat both the cosmological
constant and the density contrast at the time of recombination are
random variables. They conclude that most galaxies form at the time
of cosmological constant dominance, providing an anthropic explanation
of \eqref{eq:coinc}. The world-line measure of the previous section
provides a much simplified version of this computation, relying only
on $H(t)$ .

\subsubsection{Youngness Paradox and Oldness Paradox}

One of the more popular cosmological measures in standard eternal
inflation assigns likelihood proportional to volume of a given pocket
universe. In this picture, a region of large vacuum energy expands
very rapidly and comes to dominate the volume along a family of spacelike
hypersurfaces (see for example \cite{Guth:2007ng}. Thus for any finite
time truncation the pockets with Planck scale expansion dominate in
an overwhelming manner. This would then make our universe which has
been around for billions of years a very special old pocket \cite{Linde:1994gy}. 

The cosmological measure described in section \ref{sub:Cosmological-measure}
also resolves this paradox. A typical bubble will actually be one
with a very large number of states according to \eqref{eq:probdist}.
Since this state evolves from a singular state, an observer will see
higher density looking toward the past, but such an observer will
typically sit near the coincidence point \eqref{eq:peak}.

\subsubsection{Boltzmann Brains }

If the bubble were to be arbitrarily long-lived, it might happen that
the area under the curve in figure \ref{fig:Entropy-of-matter} associated
with Hawking radiation could dominate, versus the area under the matter/radiation
peak. This leads to a potential oldness paradox in the present scenario,
which is often referred to as the Boltzmann Brain paradox.

In the usual picture of eternal inflation expansion is eternal. This
leads to Boltzmann brain observers that form at late times from thermal
or vacuum fluctuations (see for example \cite{Dyson:2002pf}). Again
volume weighting of the expanding regions leads to an infinite number
of Boltzmann brains in comparison with ordinary observers living around
the time of cosmic coincidence. Since Boltzmann brains would infinitely
dominate over ordinary observers, observations like ours would be
atypical. Again a suitable solution to the measure problem is needed
to cure the putative catastrophe of Boltzmann brains.

In the present scenario, semiclassical physics breaks down long before
Boltzmann brains become an issue, and the problem is avoided. If the
entropy of a Boltzmann brain is $S_{BB}\gg1$ the timescale for formation
will be of order$ $\begin{equation}
\tau_{BB}=l_{dS}e^{S_{BB}}\,.\label{eq:brain}\end{equation}
This is to be compared with the timescale at which semiclassical physics
breaks down \eqref{eq:plscale}. Having a separation of scales between
$l_{dS}$ and $l_{pl}$ so that semiclassical physics is valid requires
some fine tuning, but these scales appear inside a log in \eqref{eq:plscale}.
A much higher degree of fine tuning is needed to make this time larger
than \eqref{eq:brain}. Therefore most worldlines will not see dominance
of Boltzmann brain observers.

If one does manage to find a region of the landscape where \eqref{eq:brain}
is less than \eqref{eq:plscale}, one still avoids eternal expansion
of the bubble because the worldline eventually tunnels back to the
stable AdS region as described in section \ref{sec:Consistency-checks}
on timescale given in \eqref{eq:decaytime}. We conclude that while
there may be regions of the landscape where Boltzmann brain observers
could dominate, extreme fine tuning is required. A detailed specification
of the CFT would be needed to make sharper statements.

In the scenario where an expanding bubble lives inside an exterior
SAdS geometry, there is a different type of Boltzmann brain that can
appear in the exterior region. These are potentially more dangerous.
Consider the history of such an observer who is created by a thermal
fluctuation outside the Schwarzschild black hole. Such an observer
will follow a timelike geodesic, and will enter the black hole horizon
and hit the singularity on a timescale of order $r_{bh}$. Semiclassical
gravity should be a good approximation in this exterior region, and
modulo Poincare recurrences, this region is eternal. Therefore we
expect a substantial number of such observers to appear and this number
could overwhelm the number of ordinary observers.

In this scenario, the stable AdS vacuum state is supersymmetric, so
atoms and other complex bound states will not appear. Thus the kind
of complex quasi-stable bound state corresponding to a Boltzmann brain
may simply not appear in this region. Rather such an observer would
likely need a large supersymmetry breaking bubble to surround her,
which leads us back to the original scenario.

Alternatively, for observables of interest to us, we can condition
probabilities by the requirement that observers live in an expanding
phase. This rules out this dangerous class of Boltzmann brain observers.
While it is straightforward to impose this condition in the limit
$l_{pl}\to0$ where semiclassical gravity is valid, it remains to
be seen whether this distinction is possible in the full quantum regime.

\subsubsection{Problems of time}

The time coordinate is not a physical observable in a diffeomorphism
invariant formulation of gravity. Usually the way around this is to
use a rolling scalar field to define an invariant notion of time.
However this then raises issues with recovering locality and causality.
These issues are reviewed in \cite{Isham:1992ms}.

In the AdS/CFT context, time on the conformal boundary provides a
global time coordinate. The CFT comes equipped with a well-defined
norm, and time evolution in the conformal field theory is unitarity.
The CFT norm induces a well-defined measure on the space of bubble
cosmologies, of which \eqref{eq:probdist} is a simple example. Less
clear is how bulk locality and causality emerges, but this is a topic
under active investigation.

Another problem of time is how the arrow of time arises in a CPT invariant
theory such as gravity. Let assume the history shown in figure \eqref{fig:Punctuated-Eternal-Inflation: Figure 4}
effectively explores the microcanonical ensemble in the CFT. In addition
to expanding bubble states, by CPT invariance we will also have an
equal number of time-reversed collapsing bubble states. However any
given macroscopic bubble solution that gives an expanding or contracting
region de Sitter region will necessarily be time asymmetric \cite{Lowe:2007ek}.
Thus in this scenario the arrow of time arises from spontaneous symmetry
breaking: the underlying theory is CPT invariant, but the solutions
break this symmetry.
\begin{acknowledgments}
The research of D.A.L. and S.R. is supported in part by DOE grant
DE-FG02-91ER40688-Task A. S.R. is supported in part by Lehman College,
City University of New York.
\end{acknowledgments}
\appendix

\section{Mapping global time to bubble proper time\label{sec:Appendix-A}}

In this appendix we find the relation between comoving time inside
the dS bubble, and Schwarzschild time in the AdS region. The proper
time along the bubble wall $\tau$ satisfies \eqref{eq:locals} with
the radial coordinate $r(\tau)$ determined by \eqref{eq:reom}. This
allows us to determine $t_{SAdS}(r)$ at the bubble wall via

\[
dt_{SAdS}=dr\left(\frac{1}{f_{SAdS}(r)}\left(\frac{1}{f_{SAdS}(r)}-\frac{1}{V_{eff}(r)}\right)\right)^{1/2}\,.\]

Integrating in the approximation, $r\rightarrow\infty$,\begin{equation}
t_{SAdS}(r)=B-A\frac{l_{AdS}^{2}}{r},\quad A=\sqrt{1+\frac{4\kappa^{2}l_{AdS}^{2}l_{dS}^{4}}{\left(l_{AdS}^{2}+l_{dS}^{2}(\kappa^{2}l_{AdS}^{2}-1)\right)^{2}+4l_{AdS}^{2}l_{dS}^{2}}}\label{eq:tswar}\end{equation}
with $B$ an integration constant. We see the bubble reaches $r=\infty$
at finite Schwarzschild time $t_{SAdS}=B$. 

In the interior, the $r$ coordinate becomes timelike for $r>l_{dS}$,
while $t_{dS}$ approaches a constant on the bubble wall $t_{dS}=C$.
Now let us transform to comoving time $\tau_{dS}$ in the de Sitter
patch, working with flat spatial slices

\[
ds^{2}=-d\tau_{dS}^{2}+l_{dS}^{2}e^{2\tau_{dS}}d\vec{x}^{2}\,.\]
The comoving time $\tau_{dS}$ is related to $r$ by\[
\tau_{dS}=l_{dS}\log\left(\frac{r}{l_{dS}}\right)\]
so inserting this into \eqref{eq:tswar} we obtain

\begin{equation}
t_{SAdS}=B-\frac{Al_{AdS}^{2}}{l_{dS}}e^{-\frac{\tau_{dS}}{l_{dS}}}\,.\label{eq:sadstime}\end{equation}
Now the semiclassical approximation is not expected to continue to
hold very close to the null singularity induced by back-reaction\eqref{fig:Penrose-diagrams-after}.
If we assume the semiclassical approximation holds to $t_{SAdS}=B-l_{pl}$
this translates into breakdown at comoving de Sitter time\[
\tau_{dS}=l_{dS}\log\left(\frac{Al_{AdS}^{2}}{l_{pl}l_{dS}}\right)\,.\]

\section{Modeling the bubble signature\label{sec:Appendix-B}}

In section \ref{sub:Bubble-states-in}, the signatures of the bubble
state in the CFT are described. One such signature is radiation propagating
through the bubble wall and out to infinity in the AdS region, which
can be related straightforwardly to CFT correlators. In this appendix,
we develop a simple model for this radiation signature. Working in
the semiclassical probe approximation we consider a massless (conformal)
scalar field propagating in this bubble geometry with the purely reflective
boundary conditions at the bubble wall and zero flux at the AdS spacelike
infinity. For computational convenience we specialize to $(1+1)$-d.
Since all space times are conformal to flat (Minkowski) space time
in $(1+1)$-d, this is conformally mapped to the propagation of a
scalar field in a flat background with moving mirror \cite{Birrell:1982ix}.

Introduce null coordinates $u,v$ on the flat spacetime\[
ds^{2}=dudv\,.\]
The moving mirror boundary follows the trajectory $x(u),\tau(u)$
where $u\equiv\tau(u)-x(u)$. The modes solutions are, \[
\phi_{\omega}=e^{-i\omega v}+e^{-i\omega p(u)}\]
where \[
p(u)=2\tau(u)-u.\]

In the mirror rest frame ($U$,$V$) with mirror at the origin,

\[
V=v,\qquad U=p(u)\]
or,\[
ds^{2}=dudv=\frac{1}{p'(u)}\; dUdV\,.\]
Performing a conformal transformation, the energy flux in $u,v$ frame
\cite{Birrell:1982ix}, is 

\begin{equation}
T_{uu}=\frac{1}{12\pi}\sqrt{p'}\partial_{u}^{2}\left(\frac{1}{\sqrt{p'}}\right).\label{eq:stress}\end{equation}

Now lets move on to pure AdS$_{2}$. The metric is,

\begin{eqnarray*}
ds^{2} & = & -(1+r^{2})dt^{2}+\frac{dr^{2}}{1+r^{2}},\quad t,r\in(-\infty,\infty)\\
 & = & -\sec^{2}x\:(dt^{2}-dx^{2}),\; x\equiv\tan^{-1}r\in[-\pi/2,\pi/2]\\
 & = & -\sec^{2}\frac{v-u}{2}\: dudv,\quad u,v\equiv t\mp x\,.\end{eqnarray*}

As before we now shift to a frame where the mirror is at rest at the
origin (i.e. $U=p(u)$ and $V=v$ ) and we can (conformally) relate
this to the mirror at rest in flat space case,

\begin{equation}
ds_{AdS}^{2}=-\sec^{2}\frac{v-u}{2}dudv=-\sec^{2}\left(\frac{V-p^{-1}(U)}{2}\right)\partial_{U}p^{-1}(U)\; dUdV=C(U,V)dUdV\,.\label{eq:twoads}\end{equation}
Then the renormalized energy-momentum tensor expectation in the $U,V$
vacuum is,

\[
T_{UU}=-\frac{1}{12\pi}F_{U}[C(U,V)],\qquad T_{VV}=-\frac{1}{12\pi}F_{V}(C(U,V))\]
\[
T_{UV}=\frac{RC(U,V)}{96\pi}\]
where the operator $F$ is defined as $F_{x}(y(x))=y^{1/2}\partial_{x}^{2}(y^{-1/2})$
and $R$ is the AdS$_{2}$ Ricci scalar. Now reverting to $u,v$ coordinates,\[
T_{uu}=\left(\frac{\partial U}{\partial u}\right)^{2}T_{UU}=T_{uu}^{(AdS)}+T_{uu}^{(m)}\]
where $T_{uu}^{(AdS)}=-\frac{1}{12\pi}F_{u}[\sec^{2}\frac{v-u}{2}]$
is the piece due to a static mirror and $T_{uu}^{(m)}$ due to motion
of the mirror, \begin{equation}
T_{uu}^{(m)}=-\frac{1}{12\pi}\left(\frac{dp}{du}\right)^{1/2}\frac{d}{du}\left(\frac{dp}{du}\right)^{-1/2}+\frac{1}{12\pi}\left(\frac{dp}{du}\right)^{-1/2}\sec\left(\frac{v-u}{2}\right)\frac{d}{du}\left(\frac{\frac{d}{du}\left(\frac{dp}{du}\right)^{1/2}}{\frac{dU}{du}\sec(\frac{v-u}{2})}\right)\label{eq:movmir}\end{equation}

\begin{equation}
T_{vv}^{(m)}=T_{uv}^{(m)}=0\,.\label{eq:movmirt}\end{equation}
Related results appear in \cite{Walker:1984vj}. 

Let us model the path of the expanding bubble by a trajectory with
constant acceleration $\alpha$ 

\begin{eqnarray*}
x(\tau) & = & \alpha^{-1}\cosh\alpha\tau\\
t(\tau) & = & \alpha^{-1}\sinh\alpha\tau\end{eqnarray*}
we have, \begin{equation}
u(\tau)=-\alpha^{-1}e^{-\alpha\tau},\quad v(\tau)=\alpha^{-1}e^{\alpha\tau}=-\alpha^{-2}u^{-1}(\tau)\label{eq:accel}\end{equation}
Now if we go to the rest frame of the bubble,\[
V(v)-U(u)=0,\]
Then with $V(v)=v$ and $U(u)=-\alpha^{2}u^{-1}$ the condition \eqref{eq:accel}
is satisfied. So for the uniformly accelerating mirror we have,

\[
p(u)=-\alpha^{-2}u^{-1}.\]
The energy flux at infinity can readily be checked to be finite.

The calculation can be generalized to the SAdS black hole in $1+1$-d
and the same qualitative results are obtained, though formulas are
more complicated. The stress-energy tensor splits into two different
components: Hawking radiation of the black hole; the contribution
from the reflection from the moving mirror. The reflected radiation
just alters the black hole mass as argued in the above calculation
by a finite amount. We conclude the path of the bubble may be deduced
from the radiation received at infinity in the AdS region. 

A more realistic model would derive the boundary conditions of the
fields from the detailed knowledge of the gravity theory, including
the potential of figure \ref{fig:The-potential-landscape}. In this
case the radiation reaching infinity in the AdS region will also contain
information on the internal state of the bubble, and consequently,
this information will show up CFT correlation functions in a bubble
state.

\bibliographystyle{brownphys}
\bibliography{draft}

\end{document}